\documentclass[final,3p,times,onecolumn,sort&compress]{elsarticle}

\usepackage{amssymb}
\usepackage{epstopdf}

	\usepackage{fancyheadings}
\renewcommand{\thispagestyle}[1]{} 

\begin{document}

\begin{frontmatter}



\title{The Influence of Interplanar Coupling on the Entropy and Specific Heat of the Bilayer Ferromagnet}

\author[]{Karol Sza\l{}owski\corref{label0}}
\ead{kszalowski@uni.lodz.pl}
\cortext[label0]{corresponding author}

\author[]{Tadeusz Balcerzak}

\address{Department of Solid State Physics, University of
{\L}\'{o}d\'{z}, ul. Pomorska 149/153, 90-236 {\L}\'{o}d\'{z},
Poland}

\begin{abstract}
The Pair Approximation method is applied to studies of the magnetic Ising-Heisenberg bilayer with simple cubic structure and spin $S=1/2$. The method allows to take into account quantum effects related to Heisenberg couplings. In the paper the entropy and magnetic contribution to the specific heat are calculated. A special attention is paid to the case when the planes are magnetically non-equivalent and the interplanar coupling is relatively weak. A double peak structure in the temperature dependence of the specific heat and the entropy change in the external field is found. When the system is strongly anisotropic (the case of Ising couplings) the Exact Calculations for Finite Cluster are performed for comparison with the Pair Approximation method.

\end{abstract}

\begin{keyword}
Ising model \sep anisotropic Heisenberg model \sep Ising-Heisenberg model \sep magnetic bilayer \sep  critical temperature \sep entropy \sep magnetic specific heat 
\sep magnetocaloric effect
\end{keyword}

\end{frontmatter}

\newpage

\section{Introduction}

Magnetic bilayers have attracted considerable attention in the literature \cite{Horiguchi1,Ghaemi1,Li1,Hansen1,Lipowski1,Monroe1, Mirza1,Puszkarski,Kapor,Spirin1,Qiu1, Oitmaa1, Saber1, Jascur1, Jascur2, Wiatrowski1, PhysA1,PhysA2} as these systems bridge the gap between the two- and three-dimensional magnets. Studying their properties provides some insight into the cross-over between the behaviours characteristic of various dimensionalities \cite{Araujo,Diep1}. That makes the systems particularly interesting from the theoretical point of view.\\

From the practical point of view, the low-dimensional magnets have gained immense applicational potential, which also stimulates further experimental and theoretical search \cite{Zhitomirsky1,Schmidt1,Honecker1,Honecker2,Pereira1,Honecker3,Ribeiro}. One of the possible practical application is connected with the magnetocaloric effect \cite{TishinBook,Szymczak}, providing a hope for an efficient and environmentally-friendly refrigeration technique. However, as shown in Ref.\cite{JMMM} the theoretical studies of such effect require methods which go beyond the molecular field approximation and take into account the spin-spin correlations. From this point of view it seems purposeful to develop the thermodynamical methods which are able to meet this requirement and are suitable for studies of low-dimensional magnetic systems.\\

Recently the Pair Approximation (PA) method \cite{PhysA1, PhysRev, PhysA2} has been developed and applied to the bilayer and bilayer-multilayer systems \cite{PhysA1,PhysA2}, as well as to the bulk materials with dilution \cite{PhysRev,JMMM}. The method incorporates the spin-pair correlation and is superior to molecular field approximation, allowing for the considerations of interaction anisotropy in spin space, to which molecular field approximation is completely insensitive. In particular, let us mention that PA describes the dependence of the critical temperature on interaction anisotropy and on dilution for ferromagnetic systems in a way which is in qualitative agreement with some exact estimates \cite{PhysRev}. The same applies to the dependence of the critical temperature on the effective coordination number \cite{JMMM}. The PA method can be formulated for systems with the spatial anisotropy of coupling between spins as well as for different magnetic sublattices, which opens the possibility of studying quite complex systems using this approach. In the PA the Gibbs energy can be found, hence a complete thermodynamic description can be obtained. However, as far as the bilayer system is considered, the method has been applied only for the phase diagram calculations. Other thermodynamic quantities have not been calculated up to now. In particular, for the distinct kinds of magnetic planes forming a bilayer, the magnetic entropy and the magnetic contribution to the specific heat have not been studied using this method. \\

In order to fill this gap, in the present paper we study the entropy and the  specific heat in the bilayer Ising-Heisenberg system for cases when the planes are magnetically non-equivalent, i.e., are characterized by different exchange integrals. For the sake of simplicity of the model, the planes forming a bilayer are based on simple quadratic lattice and the spin magnitude is $S=1/2$. The calculations performed cover a wide range of coupling parameters. A special attention is paid to the isothermal entropy change in the external field, as this characteristic is vital for the magnetocaloric effect description. It will be shown that some interesting phenomena occur when one plane is characterized by the Ising interactions and the other is described by the Heisenberg ones, whereas the mutual coupling of these planes is relatively weak. In such a case the temperature dependencies of the thermodynamic properties reveal a double-peak structure, which can be potentially interesting from the point of view of the magnetocaloric effect.\\

For the bilayer described by the Ising model, the exact numerical calculations for finite clusters have also been performed in order to control the PA method.\\

\section{Theoretical model}

\begin{figure}
\includegraphics[scale=0.25]{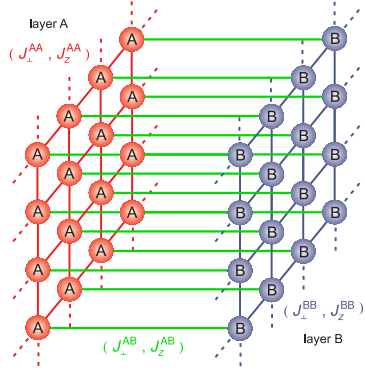}
\caption{\label{fig:fig1} A schematic view of the bilayer composed of two layers, $A$ and $B$. The intraplanar couplings are $J^{AA}_{x}=J^{AA}_{y}=J^{AA}_{\perp}$, $J^{AA}_{z}$ and $J^{BB}_{x}=J^{BB}_{y}=J^{BB}_{\perp}$, $J^{BB}_{z}$, respectively. The interplanar coupling is $J^{AB}_{x}=J^{AB}_{y}=J^{AB}_{\perp}$, $J^{AB}_{z}$. For PA calculations, the planes $A$ and $B$ are infinite with the translational symmetry for each pair. In the ECFC method a bilayer cluster consists of 4$\times$4 atoms and in-plane boundary conditions were chosen as periodic.}
\end{figure}

The bilayer consisting of two atomic planes A and B is illustrated in Fig.1. The Hamiltonian of the system in question can be written in the form of:
\begin{eqnarray}
\mathcal{H}&=&-\sum_{\left\langle i\in A,j\in A \right\rangle}^{}{\left[J_{\perp}^{AA}\left(S_{x}^{i}S_{x}^{j}+S_{y}^{i}S_{y}^{j}\right)+J_{z}^{AA}S_{z}^{i}S_{z}^{j}\right]}-\sum_{\left\langle i\in B,j\in B \right\rangle}^{}{\left[J_{\perp}^{BB}\left(S_{x}^{i}S_{x}^{j}+S_{y}^{i}S_{y}^{j}\right)+J_{z}^{BB}S_{z}^{i}S_{z}^{j}\right]}\nonumber\\
&&-\sum_{\left\langle i\in A,j\in B \right\rangle}^{}{\left[J_{\perp}^{AB}\left(S_{x}^{i}S_{x}^{j}+S_{y}^{i}S_{y}^{j}\right)+J_{z}^{AB}S_{z}^{i}S_{z}^{j}\right]}-h\sum_{i\in A}^{}{S^{i}_{z}}-h\sum_{j\in B}^{}{S^{j}_{z}}.
\label{eq1}
\end{eqnarray}
$J^{\gamma \delta}_{\perp, z}$ are the anisotropic exchange integrals and $h$ stands for the external magnetic field. We consider the case when $0\leq J^{\gamma \delta}_{x}=J^{\gamma \delta}_{y}=J^{\gamma \delta}_{\perp} \le J^{\gamma \delta}_{z}$ ($\gamma=A,B; \, \delta =A,B$), which covers all intermediate situations between the pure Ising ($J^{\gamma \delta}_{\perp}=0$) and the isotropic Heisenberg ($J^{\gamma \delta}_{\perp}=J^{\gamma \delta}_{z}$) models.\\

Following the method described in Ref. \cite{PhysA1, PhysA2} the Gibbs energy per site (i.e. the chemical potential) can be found in the Pair Approximation method. For the bilayer system with s.c. structure it can be written in the form of \cite{PhysA2}:
\begin{equation}
G=G^{AA}+G^{BB}+G^{AB}/2-2\left(G^{A}+G^{B}\right)
\label{eq2}
\end{equation}
where the single-site Gibbs energies for $A$ and $B$ planes are given by:
\begin{equation}
G^{A}=-k_{\rm B}T \,\ln\left\{2\cosh\left[\frac{\beta}{2}\left(\Lambda^{A}+h\right)\right]\right\} 
\label{eq3}
\end{equation}
and
\begin{equation}
G^{B}=-k_{\rm B}T \,\ln\left\{2\cosh\left[\frac{\beta}{2}\left(\Lambda^{B}+h\right)\right]\right\}, 
\label{eq4}
\end{equation}
respectively.
The pair Gibbs energies, where the pair spins belong to $A$ and $B$ planes, are given in the form of:
\begin{eqnarray}
G^{AA}&=&-k_{\rm B}T \,\ln \left\{2\exp\left(\frac{\beta J^{AA}_{z}}{4}\right)\cosh\left[\beta\left(\Lambda^{AA}+h\right)\right]+2\exp\left(-\frac{\beta J^{AA}_{z}}{4}\right)\cosh\left(\frac{\beta}{2}J_{\perp}^{AA}\right)\right\} 
\label{eq5}
\\
G^{BB}&=&-k_{\rm B}T \,\ln\left\{2\exp\left(\frac{\beta J^{BB}_{z}}{4}\right)\cosh\left[\beta\left(\Lambda^{BB}+h\right)\right]+2\exp\left(-\frac{\beta J^{BB}_{z}}{4}\right)\cosh\left(\frac{\beta}{2}J_{\perp}^{BB}\right)\right\} 
\label{eq6}
\\
G^{AB}&=&-k_{\rm B}T \,\ln\left\{2\exp\left(\frac{\beta J^{AB}_{z}}{4}\right)\cosh\left[\beta\left(\Lambda^{AB}+h\right)\right]+2\exp\left(-\frac{\beta J^{AB}_{z}}{4}\right)\cosh\left[\frac{\beta}{2}\sqrt{\left(\Delta^{AB}\right)^2+\left(J_{\perp}^{AB}\right)^2}\,\right]\right\} 
\label{eq7}.
\end{eqnarray}
The molecular field parameters $\Lambda^{A}$, $\Lambda^{B}$, $\Lambda^{AA}$, $\Lambda^{BB}$, $\Lambda^{AB}$ and $\Delta^{AB}$ occurring in eqs. (\ref{eq3}-\ref{eq7}) can be expressed by four variational parameters $\lambda^{AA}$, $\lambda^{BB}$, $\lambda^{AB}$ and $\lambda^{BA}$, namely:
\begin{eqnarray}
\Lambda^{A}&=&4\lambda^{AA}+\lambda^{AB}\nonumber\\
\Lambda^{B}&=&4\lambda^{BB}+\lambda^{BA}\nonumber\\
\Lambda^{AA}&=&3\lambda^{AA}+\lambda^{AB}\nonumber\\
\Lambda^{BB}&=&3\lambda^{BB}+\lambda^{BA}\nonumber\\
\Lambda^{AB}&=&2\left(\lambda^{AA}+\lambda^{BB}\right)\nonumber\\
\Delta^{AB}&=&4\left(\lambda^{AA}-\lambda^{BB}\right)
\label{eq8}
\end{eqnarray}
where $\lambda^{\gamma \delta}$ is the field acting on a spin on the sublattice $\gamma =A$ or $\gamma=B$ and originating from spins on the sublattices $\delta=A$ or $\delta =B$, respectively.\\
These four fields can be found from the variational principle for the Gibbs free energy:
\begin{equation}
\frac{\partial G}{\partial \lambda^{\gamma \delta}}=0
\label{eq9}
\end{equation}
As a result $\lambda^{\gamma \delta}$ can be found from solutions of the four variational equations:
\begin{eqnarray}
\tanh\left[\frac{\beta}{2}\left(\Lambda^{A}+h\right)\right]&=&\frac{\exp\left(\frac{\beta J^{AA}_{z}}{4}\right)\sinh\left[\beta\left(\Lambda^{AA}+h\right)\right]}{\exp\left(\frac{\beta J^{AA}_{z}}{4}\right)\cosh\left[\beta\left(\Lambda^{AA}+h\right)\right]+\exp\left(-\frac{\beta J^{AA}_{z}}{4}\right)\cosh\left(\frac{\beta J^{AA}_{\perp}}{2}\right)}
\label{eq10}
\\
\tanh\left[\frac{\beta}{2}\left(\Lambda^{A}+h\right)\right]&=&\frac{\exp\left(\frac{\beta J^{AB}_{z}}{4}\right)\sinh\left[\beta\left(\Lambda^{AB}+h\right)\right]+\frac{\Delta^{AB}}{\sqrt{\left(\Delta^{AB}\right)^2+\left(J_{\perp}^{AB}\right)^2}} \exp\left(-\frac{\beta J^{AB}_{z}}{4}\right)\sinh\left(\frac{\beta\sqrt{\left(\Delta^{AB}\right)^2+\left(J_{\perp}^{AB}\right)^2}}{2}\right)}{\exp\left(\frac{\beta J^{AB}_{z}}{4}\right)\cosh\left[\beta\left(\Lambda^{AB}+h\right)\right]+\exp\left(-\frac{\beta J^{AB}_{z}}{4}\right)\cosh\left(\frac{\beta\sqrt{\left(\Delta^{AB}\right)^2+\left(J_{\perp}^{AB}\right)^2}}{2}\right)}
\label{eq11}
\\
\tanh\left[\frac{\beta}{2}\left(\Lambda^{B}+h\right)\right]&=&\frac{\exp\left(\frac{\beta J^{BB}_{z}}{4}\right)\sinh\left[\beta\left(\Lambda^{BB}+h\right)\right]}{\exp\left(\frac{\beta J^{BB}_{z}}{4}\right)\cosh\left[\beta\left(\Lambda^{BB}+h\right)\right]+\exp\left(-\frac{\beta J^{BB}_{z}}{4}\right)\cosh\left(\frac{\beta J^{BB}_{\perp}}{2}\right)}
\label{eq12}
\\
\tanh\left[\frac{\beta}{2}\left(\Lambda^{B}+h\right)\right]&=&\frac{\exp\left(\frac{\beta J^{AB}_{z}}{4}\right)\sinh\left[\beta\left(\Lambda^{AB}+h\right)\right]-\frac{\Delta^{AB}}{\sqrt{\left(\Delta^{AB}\right)^2+\left(J_{\perp}^{AB}\right)^2}} \exp\left(-\frac{\beta J^{AB}_{z}}{4}\right)\sinh\left(\frac{\beta\sqrt{\left(\Delta^{AB}\right)^2+\left(J_{\perp}^{AB}\right)^2}}{2}\right)}{\exp\left(\frac{\beta J^{AB}_{z}}{4}\right)\cosh\left[\beta\left(\Lambda^{AB}+h\right)\right]+\exp\left(-\frac{\beta J^{AB}_{z}}{4}\right)\cosh\left(\frac{\beta\sqrt{\left(\Delta^{AB}\right)^2+\left(J_{\perp}^{AB}\right)^2}}{2}\right)}
\label{eq13}.
\end{eqnarray}
Having the variational parameters $\lambda^{\gamma \delta}$, the determination of the Gibbs energy as a function of the external field $h$ and temperature $T$ is possible. On this basis all further thermodynamic properties can be found. For instance, in this paper we focus our attention on entropy and magnetic specific heat studies. The corresponding formulas for the entropy $S$ and the specific heat $C_{h}$ are then given by:
\begin{equation}
S=-\left(\frac{\partial G}{\partial T}\right)_{h}
\label{eq14}
\end{equation}
and 
\begin{equation}
C_{h}=-T\left(\frac{\partial^2 G}{\partial T^2}\right)_{h}.
\label{eq15}
\end{equation}
The above temperature derivatives can be easily calculated numerically for constant external field $h$.\\

Knowing the temperature and external field dependence of the magnetic entropy, the further thermodynamic quantities important for the characterization of the magnetocaloric effect can be calculated \cite{TishinBook,Szymczak,JMMM}. One of them is a magnitude $|\Delta S_{T}|$ of the entropy change under the isothermal magnetization/demagnetization of the system between the external field values $h=0$ and $h>0$. It is defined by $|\Delta S_{T}|=-\int_{0}^{h}{\left(\partial m / \partial T \right)_{h'}\,dh'}$. This quantity is one of the crucial factors describing the magnetocaloric performance of the system and is commonly measured \cite{TishinBook,Szymczak,JMMM}.

In order to control the numerical calculations in the PA method we developed also another approach which is based on the Exact Calculation for Finite Clusters (ECFC). The method is applicable to the systems with Ising interactions. ECFC consists in a direct computation of all Ising states for the chosen cluster. Having the energy of each state, the statistical sum can be numerically obtained. Hence, all thermodynamic properties, including entropy and specific heat, can be calculated for arbitrary temperature. The method is sensitive to the shape of a cluster and selection of the boundary conditions. Our largest cluster consisted of $4 \times 4$ square elementary cells in each bilayer plane (in total 32 lattice sites), and the boundary conditions were chosen as periodic. The considered cluster corresponds to that illustrated in Fig.1. Let us mention that the method shares some features with Monte Carlo calculations, the difference being that here all the system states are taken into account. Therefore, some observations regarding the sensitivity of the Monte Carlo results to boundary conditions are valid here. Namely, it has been observed that periodic boundary conditions seem superior to other choices, such as free boundary conditions, for planar Ising systems \cite{Landau1,Jaan1}. Moreover, the possibility of studying larger finite clusters is severely limited by the necessary computational time, which rise exponentially with the system size for exact diagonalization methods (see e.g. \cite{Book3}), the present limit being close to 40 sites. \\
The numerical calculations based on the PA method will be presented in the next section and, whenever appropriate, compared with the ECFC approach.\\

\section{The numerical results and discussion}

\begin{figure}
\includegraphics[scale=0.25]{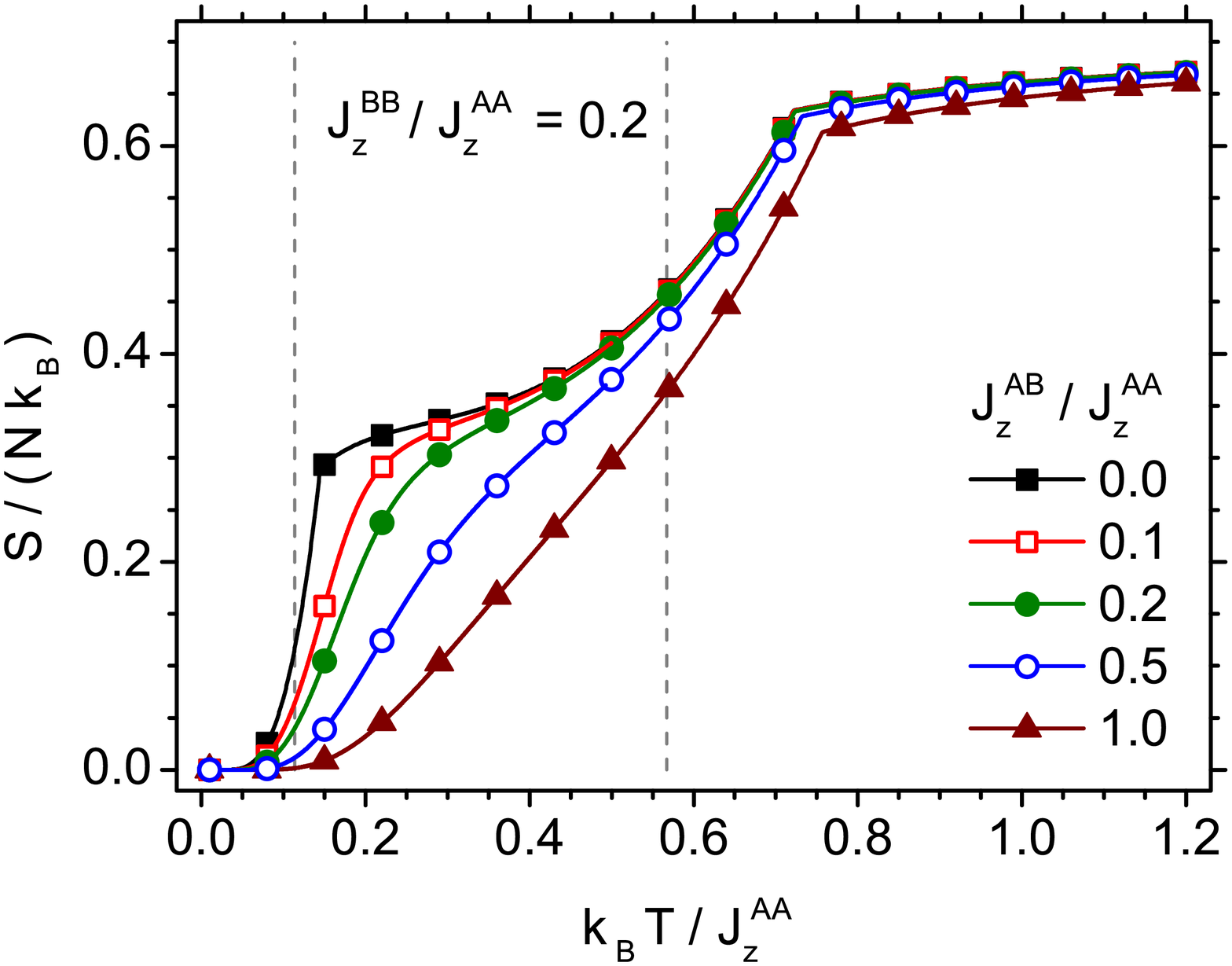}
\caption{\label{fig:fig2}The entropy per site in Boltzmann constant units vs. dimensionless temperature $k_{\rm B}T/J^{AA}_{z}$, calculated in the PA method. The bilayer is of Ising type with $J^{BB}_{z}/J^{AA}_{z}=0.2$ and $h/J^{AA}_{z}=0$. The interplanar couplings are changed from $J^{AB}_{z}/J^{AA}_{z}=0$ (i.e., when two planes are magnetically separated) until  $J^{AB}_{z}/J^{AA}_{z}=1$, when a full coupling takes place. When the planes are separated the Curie temperatures resulting from the exact Onsager solution (Ref.~\cite{Onsager1}) are indicated by the vertical dashed lines.
}
\end{figure}

The results of numerical calculations for the model developed in previous section are illustrated graphically in the plots.
In the subsequent figures the results of calculations of the entropy $S$ and magnetic specific heat $C_{h}$ as well as the magnitude of the entropy change $| \Delta S_T|$ between the external field $h$ and $0$ are presented vs. temperature for various interaction parameters $J^{\gamma \delta}_{\perp, z}$. The values of exchange integrals and external fields are normalized to $J^{AA}_{z}$, whereas the calculated thermodynamic properties are expressed  in Boltzmann constant $k_{\rm B}$ units per site. In the present paper we consider the general case when the planes are magnetically inequivalent, i.e., the interaction of spins within $A$-plane is different from that within $B$-plane.\\

As the first case we consider the pure Ising bilayer when $J^{BB}_{z}/J^{AA}_{z}=0.2$ and $h/J^{AA}_{z}=0$ (Figs.2-5). The interplanar couplings vary from $J^{AB}_{z}/J^{AA}_{z}=0$ (i.e., when the two planes are magnetically separated) until $J^{AB}_{z}/J^{AA}_{z}=1$, when considerably strong coupling takes place. The Curie temperatures for separated A and B planes resulting from the exact Onsager solution (Ref. \cite{Onsager1}) are indicated by the vertical dashed lines. The values of these temperatures are $k_{\rm B}T_{c}/J_{z}^{AA}=1/\left[2\log\left(1+\sqrt{2}\right)\right]\simeq 0.5673$ and $k_{\rm B}T_{c}/J_{z}^{AA}=1/\left[10\log\left(1+\sqrt{2}\right)\right]\simeq 0.1135$, respectively, where $J^{z}_{BB}/J^{z}_{AA}=0.2$. The PA method used here in the case of two separated planes gives the values: $k_{\rm B}T_{c}/J_{z}^{AA}=1/\left(2\log 2\right)\simeq 0.7213$ and $k_{\rm B}T_{c}/J_{z}^{AA}=1/\left(10\log 2\right)\simeq 0.1443$, respectively. (A wider discussion of the phase transition temperatures and comparison of PA with various approaches are included, for instance, in Fig.2 of Ref.~\cite{JMMM}).

As the first example in Fig.2, the entropy is shown vs. temperature as obtained in the PA method. From different curves corresponding to various  $J^{AB}_{z}$ parameters we can observe the influence of the interplanar coupling on the entropy behaviour. For instance, for the uncoupled planes two kinks of the entropy are seen, which correspond to two different phase transition temperatures of  $A$ and $B$ planes. With increase of the $A-B$ coupling the low-temperature kink vanishes and the entropy is gradually lowered, while the second kink corresponds to the common critical temperature of the whole system and is slightly shifted toward higher temperatures. The most remarkable changes are observed in the temperature region below the higher kink.\\

\begin{figure}
\includegraphics[scale=0.25]{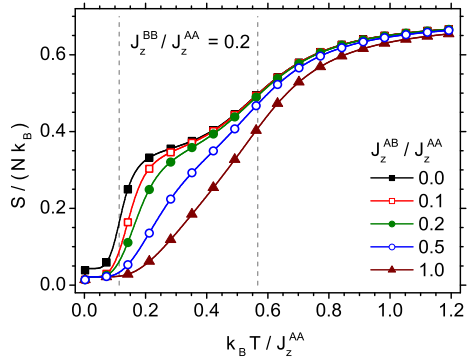}
\caption{\label{fig:fig3}The entropy per site in Boltzmann constant units vs. dimensionless temperature $k_{\rm B}T/J^{AA}_{z}$, calculated in the ECFC method. The interaction parameters are the same as in Fig.2. When the planes are separated the Curie temperatures resulting from the exact Onsager solution (Ref.~\cite{Onsager1}) are indicated by the vertical dashed lines.}
\end{figure}

For comparison, in Fig.3 the entropy obtained with the ECFC method for the same Hamiltonian parameters is presented. A qualitative agreement between results of both methods can be noted. Note that, since we perform ECFC calculations for a system composed of a finite number of $N$ spins, additional degeneracies in the ground state are present, which is not the case in the thermodynamic limit. It can be shown that they give rise to a residual entropy of $S\left(T=0\right)/N=k_{\rm B}\ln 2/N$, so that the temperature dependences of entropy do not start form the zero value in the plot. The residual entropy for $J^{AB}_{z}=0$ is twice higher than in the remaining cases because we deal  with two uncoupled systems, each containing $N/2$ lattice sites. \\

In both methods we obtain the temperature dependences of entropy which differ significantly from the molecular field theory results, which predict constant, saturated entropy of $S/N=k_{\rm B}\ln 2$ above the Curie point. This is an important qualitative improvement, which is crucial, for example, for magnetocaloric considerations \cite{JMMM}.

\begin{figure}
\includegraphics[scale=0.25]{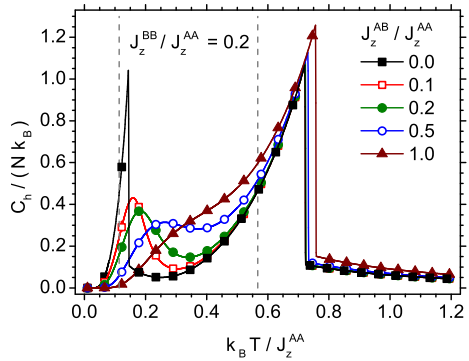}
\caption{\label{fig:fig4}The magnetic specific heat per site in Boltzmann constant units vs. dimensionless temperature $k_{\rm B}T/J^{AA}_{z}$, calculated in the PA method. The bilayer is of Ising type with the same interaction parameters as in Figs. 2 and 3. When the planes are separated the Curie temperatures resulting from the exact Onsager solution (Ref.~\cite{Onsager1}) are indicated by the vertical dashed lines.
}
\end{figure}

\begin{figure}
\includegraphics[scale=0.25]{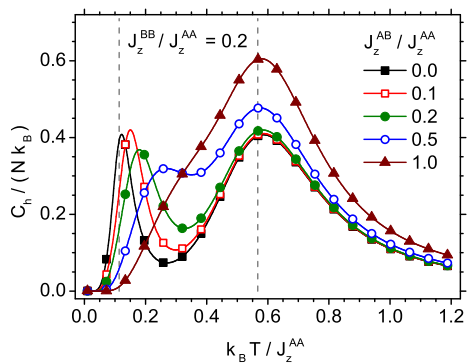}
\caption{\label{fig:fig5}The magnetic specific heat per site in Boltzmann constant units vs. dimensionless temperature $k_{\rm B}T/J^{AA}_{z}$, calculated in the ECFC method. The bilayer is of Ising type with the same interaction parameters as in Figs. 2, 3 and 4. When the planes are separated the Curie temperatures resulting from the exact Onsager solution (Ref.~\cite{Onsager1}) are indicated by the vertical dashed lines.}
\end{figure}

In Fig.4, the magnetic specific heat calculated in the PA method for the same interaction parameters as in Fig.2 is plotted. In the absence of interplanar coupling, two peaks of the specific heat are seen, corresponding to two distinct phase transition temperatures of the $A$ and $B$ planes. When the coupling $J^{AB}_{z}$ increases the first peak vanishes whereas the high-temperature peak shifts slightly to higher temperatures, and also the Curie temperature increases (see \cite{PhysA2}). Let us note that for arbitrary weak coupling, the system undergoes as a whole a single phase transition at certain temperature \cite{PhysA2}. The non-zero values of the specific heat above the Curie temperature, which are due to the non-vanishing spin-pair correlations, are taken into account in the PA method. On the other hand, when the temperature tends to absolute zero the specific heat vanishes in accordance with the third law of thermodynamics. The similar behaviour of the specific heat can be observed in Fig.5 where the ECFC method is adopted for calculations. The agreement between these two independent approaches for the Ising system is an important test for the PA, since the method is further developed for the models when the Heisenberg interactions are included. Let us note that the behaviour of the specific heat for layered systems with inequivalent magnetic planes has been also a subject of interest, for example, in the Refs.~\cite{Jascur1}, \cite{Veiller1} \cite{Ferrenberg1}. In \cite{Ferrenberg1}, mostly the case of magnetic bilayer was discussed; however, each magnetic layer consisted of more than one atomic plane, contrary to our system. In the above mentioned studies, a double-peaked structure of specific heat has also been found, with higher temperature peak occurring close to the critical temperature of the systems. Moreover, it can be inferred from our results that, while increasing the strength of the interplanar coupling, the low temperature peak becomes lower and less pronounced. The effect is much more visible in the case of PA method results, while in ECFC calculations it is rather the height of the high-temperature maximum which increases with the interplanar coupling. Let us note that the ECFC calculations were performed for rather limited system size, so that the width of the specific heat peaks is considerable. The effect of lowering of the low-temperature specific heat maximum in Ising bilayer composed of two inequivalent layers as a result of increasing coupling has also been reported in the results of the Ref.~\cite{Ferrenberg1}. For stronger interplanar interactions, the peaks tend to merge into a single peak.\\

\begin{figure}
\includegraphics[scale=0.25]{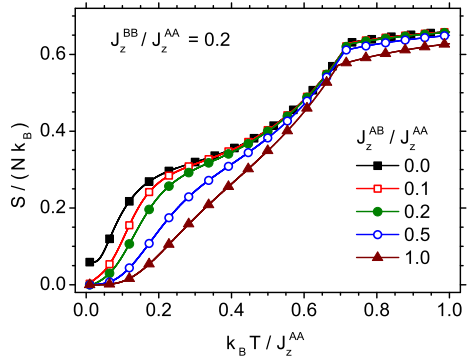}
\caption{\label{fig:fig6}The entropy per site in Boltzmann constant units vs. dimensionless temperature $k_{\rm B}T/J^{AA}_{z}$, calculated in the PA method. The $A$-$B$ bilayer is of Ising-Heisenberg type with $J^{BB}_{z}/J^{AA}_{z}=0.2$ and $h/J^{AA}_{z}=0$. The interplanar couplings are of isotropic Heisenberg-type ($J^{A B}_{\perp}=J^{A B}_{z}$) with the same strength of $J^{A B}_{z}$ as in previous figures.}
\end{figure}

\begin{figure}
\includegraphics[scale=0.25]{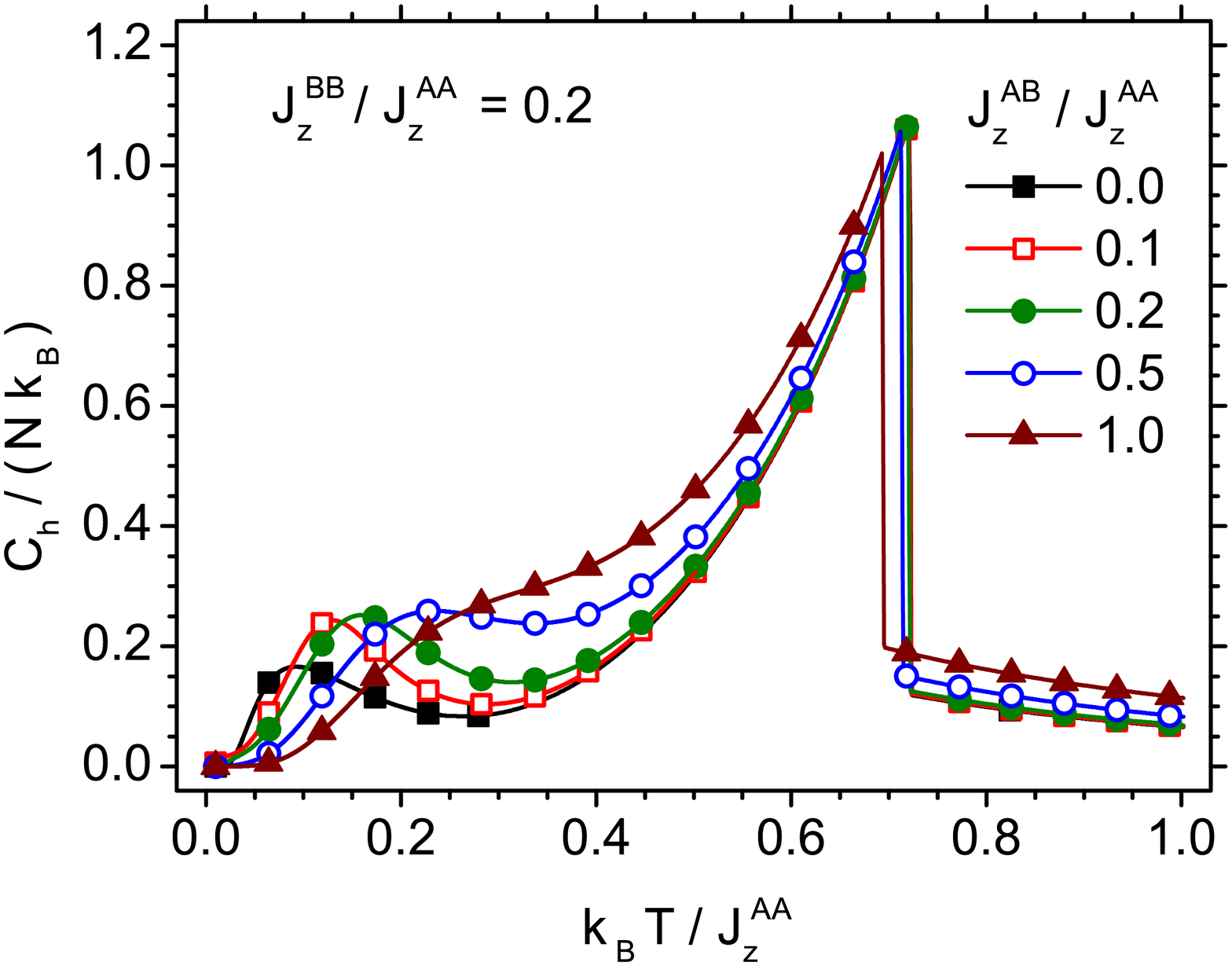}
\caption{\label{fig:fig7}The magnetic specific heat per site in Boltzmann constant units vs. dimensionless temperature $k_{\rm B}T/J^{AA}_{z}$, calculated in the PA method. The $A$-$B$ bilayer is of Ising-Heisenberg type with $J^{BB}_{z}/J^{AA}_{z}=0.2$ and $h/J^{AA}_{z}=0$. The interplanar couplings are of isotropic Heisenberg-type ($J^{A B}_{\perp}=J^{A B}_{z}$) with the same strength of $J^{A B}_{z}$ as in previous figures.}
\end{figure}

\begin{figure}
\includegraphics[scale=0.25]{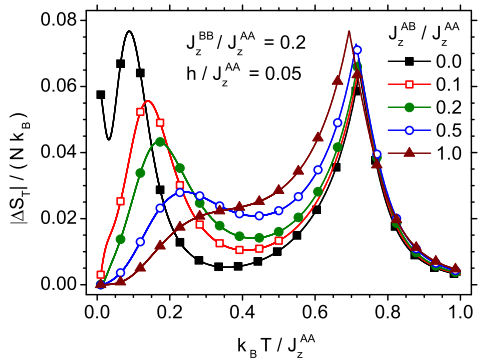}
\caption{\label{fig:fig8}The absolute value of the isothermal entropy change in the external field $h/J^{AA}_{z}=0.05$, calculated vs. dimensionless temperature in the PA method. The $A$-$B$ bilayer is of Ising-Heisenberg type with $J^{BB}_{z}/J^{AA}_{z}=0.2$. The interplanar couplings are of isotropic Heisenberg-type ($J^{A B}_{\perp}=J^{A B}_{z}$) with the same strength of $J^{A B}_{z}$ as in previous figures.
}
\end{figure}

Figs.6-8 present results obtained by means of the PA method when the interactions in magnetically weaker $B$-plane ($J^{BB}_{z}/J^{AA}_{z}=0.2$) and the interplanar $A-B$ couplings are of Heisenberg type, whereas the interactions within $A$ plane are of Ising type. Various curves correspond to the different $J^{AB}_{z}=J^{AB}_{\perp}$ coupling strength. In particular, when the planes are magnetically separated, only the Ising plane ($A$) is magnetic, since the Heisenberg plane ($B$) shows no ordering in accordance with Mermin-Wagner theorem \cite{Mermin}. By switching on the interplanar couplings, the both planes are magnetized below certain Curie temperature; however, magnetism of the Ising plane becomes slightly weakened.\\

In Fig.6 the entropy dependence on temperature is shown. An initial increase in the entropy vs. temperature for extremely weak interplanar couplings corresponds to a weak maximum of the magnetic specific heat as seen in Fig.7. In turn, this weak maximum represents a remnant of Schottky paramagnetic maximum of the specific heat, which emerges for an isolated Heisenberg plane. The entropy curves in Fig.6 show only one kink at the Curie temperature. Contrary to Figs.2-5 the Curie temperature of such Ising-Heisenberg system slightly diminishes when the interplanar coupling increases (see the effect of the Heisenberg interplanar coupling on the critical temperature of a bilayer in \cite{PhysA2}). This can be explained by disordering influence of the Heisenberg plane on the Ising one. Another characteristic feature seen in Fig.6 is the residual entropy when the interplanar couplings tend to zero. Such residual entropy originates from the isotropic Heisenberg plane at $T=0$ and is in accordance with the PA results of Ref.\cite{PhysA1}.\\

The magnetic specific heat presented in Fig.7 shows a sharp peak at the Curie temperature and a small rounded maximum at low temperatures, being the remnant of the Schottky maximum. This small maximum vanishes with increase in the interplanar coupling (again, the two maxima tend to merge into a single one for increasing $J^{AB}_{z}$). Here, Fig.7 can be compared with Fig.4 where in the case of isolated Ising planes two sharp peaks of the specific heat are seen. As the coupling between the planes increases, the principal maxima shift to opposite directions for Fig.4 and Fig.7.\\

When the external magnetic field is switched on, the system becomes more ordered and the entropy decreases. It is known that such a phenomenon is vital for the description of magnetocaloric effect \cite{JMMM}. One of its quantitative measures is the magnitude of the entropy change $\left|\Delta S_{T}\right|$ for isothermal magnetization or demagnetization between the external field of $h=0$ and $h>0$. In Fig.8 we present the absolute value of the  isothermal change of the entropy when the external field is switched on from zero value up to $h/J^{AA}_{z}=0.05$. Similarly to the specific heat, the entropy change $\left|\Delta S_{T}\right|$ also shows a double maximum behaviour. What is interesting, in the case of weak interplanar couplings, the low-temperature maximum becomes comparable with another one situated around the Curie temperature. This low temperature peak is broader, rounded and not connected with the phase transition. Because the magnitude of magnetocaloric effect is most remarkable for the maximum of $|\Delta S_{T}|$, the existence of two maxima gives also the opportunity to take into account the lower one. Therefore, we think that such a phenomenon may be of potential interest for studies of the magnetocaloric effect in bilayer systems. The main maximum tends to shift toward lower temperatures for increasing coupling. It can be also seen that for stronger couplings, i.e. for lower Curie temperatures, the maximum value achieved by $\left|\Delta S_{T}\right|$ slightly increases. Let us note that both experimental and theoretical calculations in general support the rising of  $\left|\Delta S_{T}\right|$ when the Curie temperature drops \cite{Belo1}.\\

\begin{figure}
\includegraphics[scale=0.25]{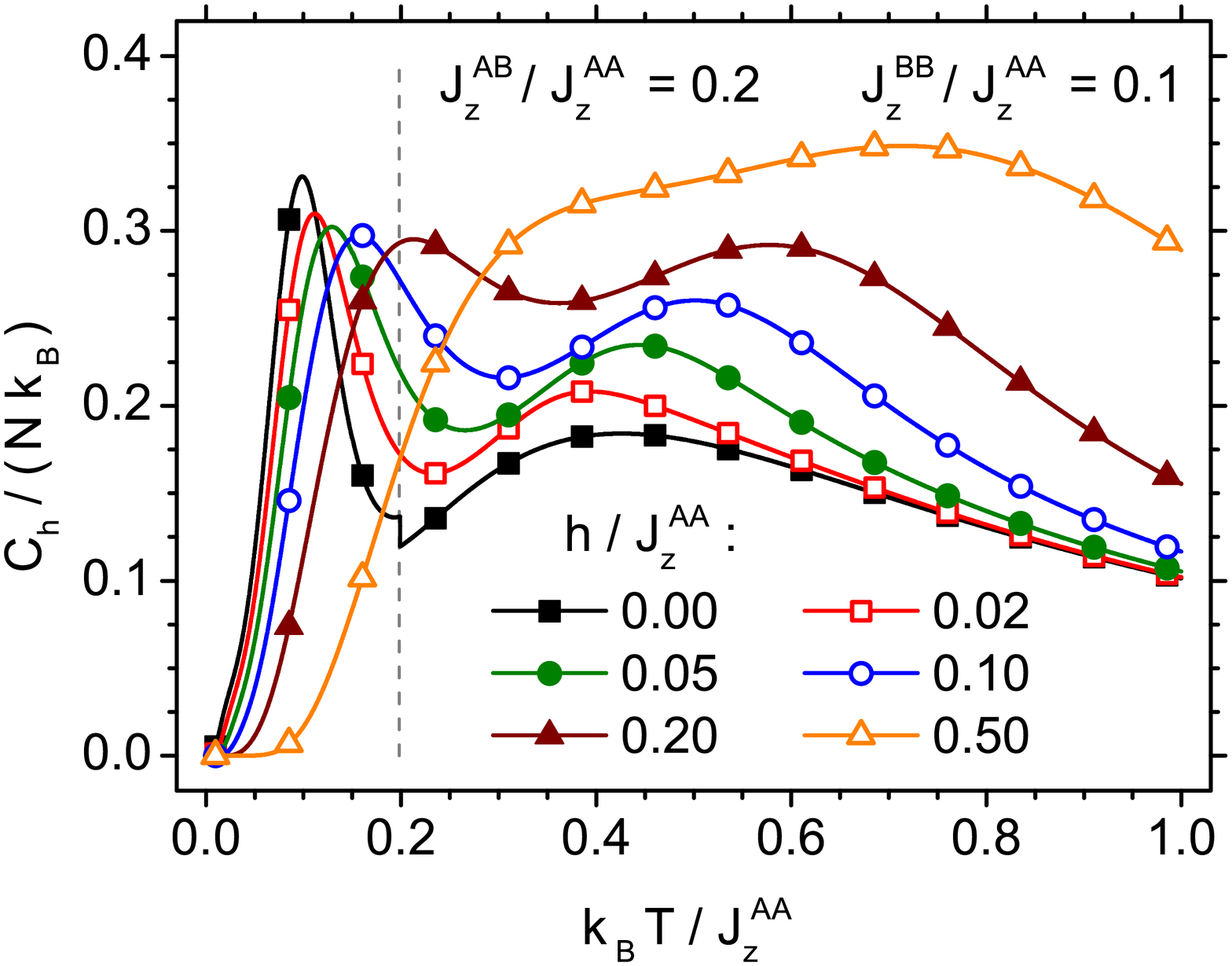}
\caption{\label{fig:fig9}The magnetic specific heat per site in Boltzmann constant units vs. dimensionless temperature $k_{\rm B}T/J^{AA}_{z}$, calculated in the PA method. The $A$-$B$ bilayer is of Heisenberg-Ising type with $J^{BB}_{z}/J^{AA}_{z}=0.1$ and $J^{A B}_{z}/J^{A A}_{z}=0.2$. The interplanar couplings are of isotropic Heisenberg-type ($J^{A B}_{\perp}=J^{A B}_{z}$). The various curves correspond to different  $h/J^{AA}_{z}$ values. By the vertical dashed line Curie temperature of the bilayer is shown.
}
\end{figure}

\begin{figure}
\includegraphics[scale=0.25]{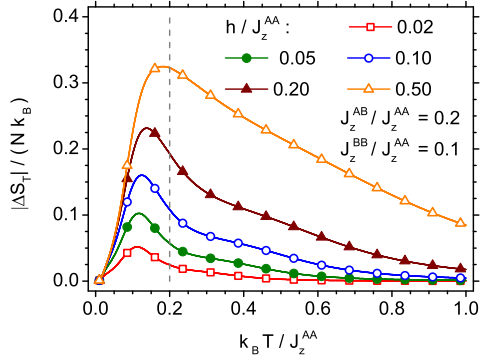}
\caption{\label{fig:fig10}Absolute value of the isothermal entropy change in the external field, calculated vs. dimensionless temperature in the PA method. The $A$-$B$ bilayer is of Heisenberg-Ising type with the same interaction parameters as in Fig.9. The various curves correspond to different  $h/J^{AA}_{z}$ values. By the vertical dashed line  Curie temperature of the bilayer is shown.}
\end{figure}

\begin{figure}
\includegraphics[scale=0.30]{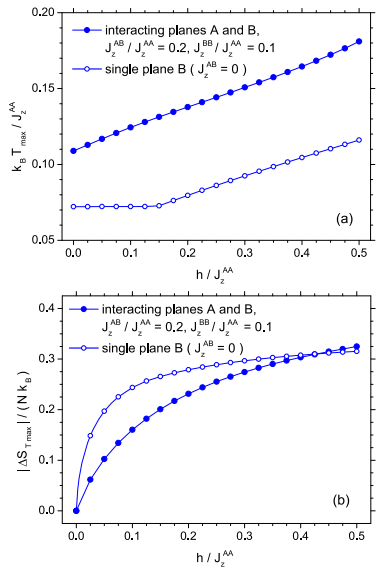}
\caption{\label{fig:fig11}(a) The characteristic temperature for which an absolute value of the isothermal entropy change in the external field reaches it maximum, as a function of the amplitude of the dimensionless external field. (b) The maximum absolute value of the isothermal entropy change in the external field, as a function of the amplitude of the dimensionless external field. Solid symbols present the results for the $A$-$B$ bilayer of Heisenberg-Ising type with the same interaction parameters as in Fig.9. Open symbols depict the data for a system where only B plane is taken into account.}
\end{figure}

In Figs. 9, 10 and 11 another interesting situation is presented when the plane $A$ with strong magnetic interaction is of Heisenberg type, whereas interactions in the $B$-plane are weak and of Ising type. The coupling between $A$ and $B$ planes is of Heisenberg type with the strength $J^{AB}_{\perp, z}/J^{AA}_{z}=0.2$. The inner interactions in both planes are characterized by the ratio $J^{BB}_{z}/J^{AA}_{z}=0.1$. The various curves plotted in Figs.9 and 10 correspond to different external fields, starting from $h/J^{AA}_{z}=0$ till $h/J^{AA}_{z}=0.5$.\\

In Fig.9 the magnetic specific heat is plotted vs. temperature. A double-maximum character of the curves is observed for small fields. When the external field increases, those two maxima merge into one broad peak. For $h=0$ the Curie temperature of the system is depicted by a vertical line. It is worth noticing that none of the peaks correspond to the Curie temperature, which lies between those two maxima. It can be explained by the fact that the higher, low-temperature peak originates from the Curie point of the single Ising ($B$) plane, whereas the high-temperature broad maximum corresponds to the Schottky anomaly of the single Heisenberg ($A$) layer. Since the two planes are coupled, the Heisenberg plane becomes magnetized and the Curie temperature of the whole system increases. The increase in the Curie temperature is due to the fact that the Heisenberg ($A$) plane, when it becomes ordered, is magnetically stronger than the Ising one. Thus, the present effect is just opposite to that reported in Fig.7, where the Curie temperature of the Ising plane was lowered as a result of coupling with the Heisenberg layer. In a view of the fact that the external field makes the magnetic ordering more uniform, the double-maximum structure vanishes in strong fields. Before it takes place, both maxima are shifted toward higher temperatures.\\

The absolute values of the isothermal entropy changes $\left|\Delta S_T\right|$ in the external fields are presented in Fig.10. In this case only one maximum for each curve is seen. The peak value of entropy change increases with increase of the field and is shifted towards higher temperatures, which effect is discussed further in detail. The shift effect is in agreement with the specific heat behaviour seen in Fig.9 and is caused by an increasing role of the Heisenberg plane. Thus, the effect is a straightforward consequence of the coupling of two planes with different magnetic characteristics. Such a fact will have further consequences for the magnetocaloric effect. The shift of the $\left|\Delta S_T\right|$ peak towards higher temperatures when the external field increases has been discussed in \cite{Franco}. It can be also observed that the observed peak is strongly asymmetric, and the fall of entropy change above the Curie temperature is rather slow, especially for higher fields.\\

The behaviour of the temperature $T_{max}$, for which the isothermal entropy change reaches a maximum is plotted in Fig.11(a) as a function of the magnetic field amplitude $h$. The data for two inequivalent coupled magnetic planes are plotted with solid symbols. It is evident that $T_{max}$ shifts approximately linearly with external magnetic field amplitude and the relative magnitude of the change can be considerable. For the purpose of comparison, $T_{max}$ is plotted also for the case when only one Ising plane (B) is present (open symbols) and the plane A does not exist. In such a situation, it is visible that for weak fields, the characteristic temperature remains constant, but for stronger fields, a linear increase is observed with a slope similar to that in the former case. Comparing the two curves, one can note that $T_{max}$ is noticeably increased when the coupling with the Heisenberg (A) plane is included.

We also plot the maximum values of isothermal entropy change $|\Delta S_{T}|$ as a function of the external field amplitude in Fig.11(b). A similar comparison like in Fig.11(a) is made. In both cases, the value of isothermal entropy change increases non-linearly with field amplitude, however, the low-field slope is considerably smaller for a system with both planes A and B present. On the contrary, for single Ising (B) plane the initial increase is much faster. Therefore, in this particular case, the interplanar coupling decreases the magnitude of $|\Delta S_{T}|$.

\section{Conclusions}

In this work we have presented the entropy and  magnetic specific heat calculations for the bilayer system. An interesting case is discussed when the exchange integrals in both planes are of unequal strength, and the interplanar coupling (Ising or Heisenberg) is relatively weak. In such a case, two peaks in the temperature dependence of the specific heat can be observed. In particular, when two Ising-Heisenberg layers are magnetically separated ($J^{A B}_{\perp, z}=0$) those peaks correspond either to the phase transition temperature (for the Ising plane), or to the paramagnetic (Schottky) maximum for the Heisenberg plane. The absolute value of the entropy change in the field may either reveal a double peak structure, similarly to the specific heat, or a single broad maximum. When the field increases, there is a shift of $|\Delta S_T|$ peak towards higher temperatures. Such a phenomenon may be of potential interest for investigation of the magnetocaloric effect.\\

In a particular case, when two planes are of Ising-type, the Exact Calculations for Finite Clusters can be done and the results are compared with the PA method. A qualitative agreement has been obtained between both methods. It can also be noted that the ECFC method yields better accuracy then PA for the Ising model, due to larger clusters considered. However, for the quantum (Heisenberg) magnets the numerical diagonalisation of the finite cluster Hamiltonian would be much less efficient, and therefore, the analytical PA method seems to be an optimum choice. \\

We think that the studies can be extended for the multilayer systems consisting of alternating planes which are magnetically non-equivalent and weakly coupled.\\

\section{Acknowledgements}

The computational support for ECFC calculations on Hugo cluster at Department of Theoretical Physics and Astrophysics, P.J. \v{S}af\'{a}rik University in Ko\v{s}ice is gratefully acknowledged.

This work has been supported by the Polish Ministry of Science and Higher Education on a special purpose grant to fund the research and development activities and tasks associated with them, serving the development of young
scientists and doctoral students.


\bibliographystyle{elsarticle-num}

\end{document}